%
%A SET OF MACROS FOR WRITING PAPERS.
%
%\today GIVES TODAY'S DATE.
%
\def\today{\ifcase\month\or January\or February\or March\or April\or May\or
June\or July\or August\or September\or October\or November\or December\fi
\space\number\day, \number\year}
%
%\note{footnote} GIVES SEQUENTIALLY NUMBERED FOOTNOTES.
%
\newcount\notenumber

\def\note{\global\advance\notenumber by 1 \footnote{$^{\the\notenumber}$}}
%
%\numbereq SEQUENTIALLY NUMBERS EQUATIONS ON THE RIGHT (number)
%
\newif\ifsectionnumbering
\newcount\eqnumber
\def\cleareqnumber{\eqnumber=0}
\def\numbereq{\global\advance\eqnumber by 1
\ifsectionnumbering\eqno(\the\secnumber.\the\eqnumber)\else\eqno
(\the\eqnumber) \fi}
\def\eqalinno{{\global\advance\eqnumber by 1}
\ifsectionnumbering(\the\secnumber.\the\eqnumber)\else(\the\eqnumber)\fi}
\def\name#1{\ifsectionnumbering\xdef#1{\the\secnumber.\the\eqnumber}
\else\xdef#1{\the\eqnumber}\fi}

\sectionnumberingtrue
%
%\ref{\name} GIVES SEQUENTIALLY NUMBERED
%REFERENCES [number], AND ASSIGNS
%THAT NUMBER TO A MACRO \name AND WRITES REF. TO FILE 1.
%
\newcount\refnumber

\immediate\openout1=refs.tex
\immediate\write1{\noexpand\frenchspacing}
\immediate\write1{\parskip=0pt}
\def\ref#1#2{\global\advance\refnumber by 1%
[\the\refnumber]\xdef#1{\the\refnumber}%
\immediate\write1{\noexpand\item{[#1]}#2}}
\def\tie{\noexpand~}

%
% NEW SECTION: \newsection The Method. (terminate with a .)
%
\font\twelvebf=cmbx10 scaled \magstep1
\newcount\secnumber

\def\newsection#1.{\ifsectionnumbering\cleareqnumber\else\fi%
	\global\advance\secnumber by 1%
	\bigbreak\bigskip\par%
	\line{\twelvebf \the\secnumber. #1.\hfil}\nobreak\medskip\par\noindent}
%
%
%\Box GIVES WAVE OPERATOR, OR LAPLACIAN
%
\def \sqr#1#2{{\vcenter{\vbox{\hrule height.#2pt
	\hbox{\vrule width.#2pt height#1pt \kern#1pt
		\vrule width.#2pt}
		\hrule height.#2pt}}}}

%
%
%\twocolumns GIVES TWO-COLUMN OUTPUT
%
\newdimen\fullhsize
\def\fiddle{\fullhsize=6.5truein \hsize=3.2truein}
\def\fullline{\hbox to\fullhsize}
\def\mkhdline{\vbox to 0pt{\vskip-22.5pt
	\fullline{\vbox to8.5pt{}\the\headline}\vss}\nointerlineskip}
\def\mkftline{\baselineskip=24pt\fullline{\the\footline}}
\let\lr=L \newbox\leftcolumn
\def\twocolumns{\fiddle
	\output={\if L\lr \global\setbox\leftcolumn=\columnbox
		\global\let\lr=R \else \doubleformat \global\let\lr=L\fi
		\ifnum\outputpenalty>-20000 \else\dosupereject\fi}}
\def\doubleformat{\shipout\vbox{\mkhdline
		\fullline{\box\leftcolumn\hfil\columnbox}
		\mkftline} \advancepageno}
\def\columnbox{\leftline{\pagebody}}
%\nosectionnumbering
\magnification=1200
\def\pr#1 {Phys. Rev. {\bf D#1\tie }}
\def\pe#1 {Phys. Rev. {\bf #1\tie}}
\def\pre#1 {Phys. Rep. {\bf #1\tie}}
\def\pl#1 {Phys. Lett. {\bf #1B\tie }}
\def\prl#1 {Phys. Rev. Lett. {\bf #1\tie }}
\def\np#1 {Nucl. Phys. {\bf B#1\tie }}
\def\ap#1 {Ann. Phys. (NY) {\bf #1\tie }}
\def\cmp#1 {Commun. Math. Phys. {\bf #1\tie }}
\def\imp#1 {Int. Jour. Mod. Phys. {\bf A#1\tie }}
\def\mpl#1 {Mod. Phys. Lett. {\bf A#1\tie}}
\def\jhep#1 {JHEP {\bf #1\tie}}
\def\nuo#1 {Nuovo Cimento {\bf B#1\tie}}
\def\ussr#1 {Zh. Eksp. Teor. Fiz {\bf #1\tie }}
\def\jetp#1 {Soviet Physics JETP {\bf #1\tie }}
\def\zetf#1 {Zh. Eksp. Teor. Fiz. {\bf #1\tie }}
\def\rmp#1 {Rev. Mod. Phys. {\bf #1\tie }}
\def\tie{\noexpand~}
\def\ov{\bar}

\parskip=15pt plus 4pt minus 3pt
\headline{\ifnum \pageno>1\it\hfil The Ginzburg-Landau theory and
the surface energy $\ldots$\else \hfil\fi}
\font\title=cmbx10 scaled\magstep1
\font\tit=cmti10 scaled\magstep1
\footline{\ifnum \pageno>1 \hfil \folio \hfil \else
\hfil\fi}
\raggedbottom

%%%%%%%%%%%%%%%%%%%%%%%
% jtl definitions

\overfullrule0pt

%%%%%%%%%%%%%%%%%%%%%%%

\rightline{\vbox{\hbox{RU03-01-B}\hbox{hep-th/0305235}}}
\vfill
\centerline{\title THE GINZBURG-LANDAU THEORY AND THE SURFACE ENERGY}
\centerline{\title  OF A COLOUR SUPERCONDUCTOR}
\vfill
{\centerline{\title Ioannis Giannakis${}^{a}$
and Hai-cang Ren${}^{a}$ \footnote{$^{\dag}$}
{\rm e-mail: \vtop{\baselineskip12pt
\hbox{giannak@summit.rockefeller.edu, ren@summit.rockefeller.edu,}}}}
}
\medskip
\centerline{$^{(a)}${\tit Physics Department, The Rockefeller
University}}
\centerline{\tit 1230 York Avenue, New York, NY
10021-6399}
\vfill
\centerline{\title Abstract}
\bigskip
{\narrower\narrower
We apply the Ginzburg-Landau theory to the colour superconducting
phase of a lump of dense quark matter. We calculate
the surface energy of a domain wall separating the normal phase from
the super phase with the bulk equilibrium maintained 
by a critical external magnetic field. Because of the
symmetry of the problem, we are able to simplify
the Ginzburg-Landau equations and express them in
terms of two components of the di-quark condensate
and one component of the gauge potential. 
The equations also contain two
dimensionless parameters: the Ginzburg-Landau parameter ${\kappa}$
and ${\rho}$. The main result of this paper is a set of
inequalities obeyed by the critical value of the Ginzburg-Landau
parameter--the value of ${\kappa}$ for which the surface energy
changes sign--and its derivative with respect to
${\rho}$. In addition we prove a number of inequalities of
the functional dependence of the surface energy on the
parameters of the problem and obtain a numerical
solution of the Ginzburg-Landau equations. Finally
a criterion for the types of colour superconductivity (type I or
type II) is established in the weak coupling approximation.
\par}
\vfill\vfill\break

%%%%%%%%%%%%%%%%%%%%%%%

\newsection Introduction.%

The Ginzburg-Landau theory provides a powerful tool for exploring
systems with inhomogeneous order parameters near the
critical temperature. For instance, its application to the surface
energy of a normal-superconducting interface
led to the discovery of the
two types of superconductors. When an external magnetic field 
whose magnitude is less than
a critical value $H_c$,
is applied to type I superconductors, the field is
expelled from the interior of
the sample (the Meissner effect). As the magnitude of the magnetic field 
increases above $H_c$,
superconductivity disappears completely and the normal
phase is restored. In the case of
type II superconductors, the
sample behaves in a similar manner as long as the
external magnetic field remains below a lower critical magnitude,
$H_{c1}$. A qualitatively different behaviour
emerges when the applied magnetic field
lies in the range $H_{c1}<H_{c}<H_{c2}$, where it becomes energetically 
favourable for the magnetic field
to penetrate the sample in the form of quantized flux lines called vortices.
Beyond the upper critical magnitude of the
field, the vortices are too dense to maintain the
condensate, and the normal phase is once again restored.

The criterion that determines
whether a superconductor is of type I or type II is
the Ginzburg-Landau parameter. It is defined as
the ratio of the penetration depth of the magnetic field $\delta$
over the coherence length $\xi$, the distance
over which changes in the order parameter occur,
i. e.
$\kappa={{\delta}\over {\xi}}$.
These characteristic lengths imply that fluctuations in the magnitude
of the condensate decay as $e^{{-\sqrt{2}}{x\over {\xi}}}$ while the
magnetic field inside the superconductor falls off as
$e^{-{x\over {\delta}}}$. In their original paper, Ginzburg and
Landau \ref{\landau}{V. L. Ginzburg and L. D. Landau, \zetf20,
1064 (1950).}
analyzed the energy of the interface between a normal
and a superconducting phase, kept in equilibrium in the bulk by an external
magnetic field at the critical value. They found analytically
that the surface energy vanishes at
$$
\kappa=\kappa_c={1\over\sqrt{2}}\cong 0.707.
\numbereq\name{\eqpene}
$$
The physical meaning of
this critical value was clarified further by Abrikosov
\ref{\abri}{A. A. Abrikosov, \jetp5, 1174 (1957).}.
It represents the demarcation line between type I
$(\kappa < {1\over\sqrt{2}})$ and type II
$(\kappa > {1\over\sqrt{2}})$ superconductors.

In the present paper we shall carry out a similar analysis of the
Ginzburg-Landau theory for colour superconductors.
Colour superconductivity is essentially
the quark analog of BCS superconductivity \ref{\col}
{B. Barrois, \np129 (1977) 390;
S. Frautschi, {\it Proceedings of the Workshop on Hadronic Matter at
Extreme Energy Density}, N. Cabibbo, ed., Erice, Italy (1978);
D. Bailin and A. Love, \pre107 (1984) 325, and references
therein for early works.}, \ref{\alf}
{M. Alford, K. Rajagopal and F. Wilczek, \np537, 443 (1999).}, 
\ref{\rssv}{R. Rapp, T. Schafter, E. V. Shuryak, M. Velkovsky, 
\prl81, 53 (1998).}, \ref{\review}{K. Rajagopal and F. Wilczek, in 
{\it At the Frontier of Particle Physics/Handbook of QCD}, B. L. Ioffe 
Festschrift, edited by M. Shifman (World Scientic, Singapore);
T. Schafer, Quark Matter, hep-ph/0304281.}.
Because of attractive quark-quark
interactions in QCD, the Fermi sphere of quarks becomes unstable
against the formation of Cooper pairs and the system becomes
superconducting at sufficiently low temperatures.
In this paper
we shall consider a domain wall separating the normal and
colour superconducting phases in equilibrium 
under the influence of an
external magnetic field and calculate the surface free energy
per unit area. The Ginzburg-Landau equations for this system 
are considerably more complicated than the analogous equations
for an ordinary superconductor \ref{\baym}{K. Iida 
and G. Baym, \pr63, 074018 (2001).}, \ref{\ren}
{I. Giannakis and H-c. Ren, \pr65, 054017 (2002).},
\ref{\Iida}{K. Iida and
G. Baym, \pr65, 014022 (2002); \pr66, 014015 (2002).}, 
\ref{\nsy}{E. Nakano, T. Suzuki and H. Yabu, J. Phys. G29, 491 (2003).},
\ref{\blaschke}{D. Blaschke and D. Sedrakian, Ginzburg-Landau equations 
for superconducting quark matter in neutron stars, nucl-th/0006038.},
because of the non-abelian nature of the problem.
By using symmetry arguments, we were able
to simplify the
problem of the domain wall 
and express the G-L equations in terms of one gauge potential and 
two components of the di-quark condensate. When expressed in terms of 
dimensionless quantities, the G-L equations
contains two dimensionless parameters:
the usual Ginzburg-Landau parameter $\kappa$ and another parameter $\rho$.
Weak coupling calculations fix the value of $\rho$
to be equal to $-{1\over 2}$.
Although we were unable to derive an analytical result analogous
to (\eqpene) we proved the following inequalities
for the critical G-L parameter and its derivative with
respect to $\rho$:
$$
\kappa_c(\rho) \le {1\over {\sqrt 2}}, \qquad
{{d{\kappa_c}}\over {d\rho}} \ge 0.
\numbereq\name{\eqvera}
$$
We also derived a number of inequalities
of the functional dependence of the surface energy 
on the parameters of the problem
and obtained a numerical solution for $\kappa_c(\rho)$. For weak coupling
we found that
$$
\kappa_c\simeq 0.589,
\numbereq\name{\eqpenecsc}
$$
in contrast to (\eqpene).

In section 2 of this paper, we shall review the
Ginzburg-Landau theory of colour
superconductivity.
The analytical and numerical solution of the domain wall 
problem will be presented in section 3. In the final section
of the paper we shall summarize our results and in the appendix we
shall discuss the validity of the Ginzburg-Landau
theory in the presence of
the fluctuations of the gauge field.

\newsection The Ginzburg-Landau Theory of Colour Superconductivity.%

The symmetry group of QCD in the chiral limit is 
$$
SU(3)_c\times SU(3)_{f_R}\times SU(3)_{f_L}\times U(1)_B,
\numbereq\name{\eqsymmetry},
$$
where the subscript $c$ denotes colour, the
subscript $f_R(f_L)$ the
right(left)-hand flavour and $B$ the baryon number. 
The electromagnetic gauge group $U(1)_{\rm em}$ is
not a separate symmetry
group, but a subgroup of $SU(3)_{f_R}\times SU(3)_{f_L}$. 
The dominant pairing channel
consists of two quarks of the same helicity and the corresponding 
order parameters in a colour superconducting quark matter will be 
denoted by ${\Psi_R}$ and ${\Psi_L}$. These condensates
transform into each
other under space inversions. Neglecting the
parity violating processess,
the Ginzburg-Landau free energy functional which is
consistent with the symmetry group (\eqsymmetry) reads
$$
\eqalign{
\Gamma=\int d^3{\vec r}\lbrace{1\over 4}F_{ij}^lF_{ij}^l
+{1\over 2}({\vec{\nabla}}\times{\vec{\cal A}})^2&+
{1\over 2}{\rm{Tr}}[({\vec D}\Psi_R)^{\dag}({\vec D}\Psi_R)
+({\vec D}\Psi_L)^{\dag}({\vec D}\Psi_L)] \cr
&+{1\over 2}a{\rm{Tr}}[\Psi_R^{\dag}\Psi_R+\Psi_L^{\dag}\Psi_L] 
+{1\over 4}b{\rm{Tr}}[(\Psi_R^{\dag}\Psi_R)^2+(\Psi_L^{\dag}\Psi_L)^2] \cr 
&+{1\over 4}b^\prime[({\rm{Tr}}\Psi_R^{\dag}\Psi_R)^2
+({\rm{Tr}}\Psi_L^{\dag}\Psi_L)^2] \cr
&+{1\over 2}c{\rm{Tr}}
(\Psi_R^{\dag}\Psi_R){\rm{Tr}}(\Psi_L^{\dag}\Psi_L)]
\rbrace \cr}
\numbereq\name{\eqrseaa}
$$
where the gauge covariant derivative of
the di-quark condensate $\Psi$ reads
$$
\eqalign{
({\vec D}{\Psi_{R(L)}})^{c_1c_2}_{f_1f_2}=&{\vec\nabla}
(\Psi_{R(L)})^{c_1c_2}_{f_1f_2}-ig{\vec A}^{c_1c'}
(\Psi_{R(L)})^{c'c_2}_{f_1f_2}-ig{\vec A}^{c_2c'}
(\Psi_{R(L)})^{c_1c'}_{f_1f_2}\cr
&-ie(q_{f_1}+q_{f_2})
{\vec{\cal A}}(\Psi_{R(L)})^{c_1c_2}_{f_1f_2}\cr}
\numbereq\name{\eqnatali}
$$
where ${\vec A}={\vec A}^l T^l$ denotes the classical vector
potential of the $SU(3)$ colour gauge field, ${\vec{\cal A}}$
the electromagnetic field, and $eq_f$ is the electric charge of the $f$-th
flavour quark. The $SU(3)$ generator $T^l$ is in its fundamental 
representation.

At weak coupling, the parameters are calculable from either
the perturbative one-gluon exchange interaction of QCD or from 
the Nambu-Jona-Lasinio effective action. Both approaches lead 
to the same expression [\baym], [\ren]:
$$
\eqalign{
a&={48\pi^2\over 7\zeta(3)}k_B^2T_c(T-T_c),\cr
b&={576\pi^4\over 7\zeta(3)}\Big({k_BT_c\over\mu}\Big)^2, \cr
b^\prime&=c=0, \cr}
\numbereq\name{\eqcoefffb}
$$
where $\mu$ is the chemical potential and $T_c$ the transition temperature.

Since in this paper we address the domain wall problem-
the interface between a normal and a superconducting phase in an external
magnetic field-
the boundary conditions select the even parity sector of (\eqrseaa)
$$
\Psi_R=\Psi_L\equiv\Psi.
\numbereq\name{\eqparity}
$$

The one-gluon exchange process that dominates the di-quark interaction 
at ultrahigh chemical potential is attractive for
quarks within the colour
antisymmetric channel. Thus the Cooper pairs realise the colour 
antisymmetric representation to the leading order of the QCD running 
coupling constant at $T=0$ \ref{\sch}{T. Schafer, \np575,
(2000) 269.}, \ref{\sho}{I. Shovkovy and L.C. R. Wijewardhana,
\pl470 (1998) 189.}, or to the leading order of
$(1-{T\over T_c})$ near $T_c$ [\ren].
Assuming such a pairing pattern persists even at
moderately high chemical potential, the
di-quark condensate $\Psi$ can be expressed in terms of a
$3\times 3$ complex
matrix, $\Phi$
$$
{\Psi}^{c_1 c_2}_{f_1 f_2}
=\epsilon^{c_1 c_2 c}\epsilon_{f_1 f_2 f}\Phi^{c}_{f}.
\numbereq\name{\eqnatal}
$$
The gauge covariant derivative of $\Phi$ can be written as
$$
({\vec D}\Phi)_f^c={\vec\nabla}\Phi_f^c-ig{\vec{\ov A}}^{cc^\prime}
\Phi_f^{c^\prime}-ieQ_f{\vec{\cal A}}\Phi_f^c
\numbereq\name{\eqmnata}
$$
where
$$
{\vec{\ov A}}={\vec A}^l{\ov T}^l, \qquad \Phi=\Phi_0+
{\Phi^l}{\ov T}^l
\numbereq\name{\eqrioh}
$$
with ${\ov T}^l=
-T^{l*}$ being the generator of the ${\bf {\ov 3}}$ representation,
$Q_f=q_{f_1}+q_{f_2}$ and $ff_1f_2$ represent a cyclic
permutation of $1, 2, 3$. Arranging the flavour index in the conventional
order of $u, d, s$ we find the diagonal electric charge matrix
$$
Q=-diag ({2\over 3}, -{1\over 3}, -{1\over 3})=
-{2\over {\sqrt 3}}{\ov T}^8.
\numbereq\name{\eqoktw}
$$
Here we have adapted an expression of $T^l$ that differs
from the standard one by a cyclic permutation of rows (columns),
in which ${\ov T}^8={1\over {2\sqrt 3}}diag (2, -1, -1)$.
The equation (\eqmnata) takes the matrix form
$$
{\vec D}\Phi={\vec\nabla}\Phi-ig{\vec{\ov A}}
\Phi+i{2\over {\sqrt 3}}e{\vec{\cal A}}\Phi{\ov T}^8,
\numbereq\name{\eqmnatata}
$$
and the Ginzburg-Landau free energy (\eqrseaa)
becomes
$$
\Gamma=\int d^3{\vec r}\Big[{1\over 4}F_{ij}^lF_{ij}^l
+{1\over 2}({\vec{\nabla}}\times{\vec{\cal A}})^2+
4{\rm{tr}}({\vec D}\Phi)^{\dag}({\vec D}\Phi)
+4a{\rm{tr}}\Phi^{\dag}\Phi+b_1{\rm{tr}}(\Phi^{\dag}\Phi)^2 
+b_2({\rm{tr}}\Phi^{\dag}\Phi)^2\Big]
\numbereq\name{\eqrsebb}
$$
where the parameters $b_1$ and $b_2$ are related to the
parameters of equation (\eqrseaa) in the following manner
$$
b_1=b, \qquad b_2=b+8b^{\prime}+8c.
\numbereq\name{\eqeric}
$$

First, let us review the case with a homogeneous condensate, i.e. 
$\vec A=\vec {\cal A}=\vec\nabla\Phi=0$, for which
the Ginzburg-Landau free energy (\eqrsebb) becomes
$$
\Gamma=\int d^3{\vec r}\Big[
4a{\rm{tr}}\Phi^{\dag}\Phi+b_1{\rm{tr}}(\Phi^{\dag}\Phi)^2 
+b_2({\rm{tr}}\Phi^{\dag}\Phi)^2\Big].
\numbereq\name{\eqrsebe}
$$
We distinguish the following three regions of the
parameter space $b_1-b_2$,
following the treatment of [\baym]
\noindent

1) $b_1>0$ and $b_1+3b_2>0$:
The minimum free energy corresponds to the
colour-flavour locked condensate [\alf],
$$
\Phi={\phi}_{0}U
\numbereq\name{\eqcfl}
$$
where
$$
\phi_0=\sqrt{-{2a\over b_1+3b_2}}
\numbereq\name{\eqwer}
$$
and $U$ is a unitary matrix.\footnote{$^{\dag}$}
{\rm It is very important to maintain this general form of the CFL 
condensate, since its nontrivial winding onto the gauge group, 
$SU(3)_c\times U(1)_{\rm em}$ gives rise to vortex filaments.} 
Consequently, we find that
$$
\Gamma_{\rm min}=-\Omega{12a^2\over b_1+3b_2}.
\numbereq\name{\eqfmin}
$$
2) $b_1<0$ but $b_1+b_2>0$: In this case the 
colour-flavour locked condensate
(\eqcfl) becomes a saddle point of the free energy, which
is neverthless
bounded from below. The minimum corresponds to an
isoscalar condensate, given
by
$$
\Phi={\rm diag}({\phi^{\prime}}_{0}
e^{i\alpha},0,0).
\numbereq\name{\eqisosclr}
$$ 
where ${\phi^\prime}_0=\sqrt{-{2a\over b_1+b_2}}$. We find then that
$$
\Gamma_{\rm min}=-\Omega{4a^2\over b_1+3b_2}.
\numbereq\name{\eqfminis}
$$
3) For $b_1$ and $b_2$ outside the region specified by 1) and 2), the free 
energy is no longer bounded from below.
Higher powers of the order parameter have
to be included and the superconducting transition 
becomes first order.
Since case 2) is mathematically identical to a metallic superconductor, 
we shall not address it in this paper.

In order to identify the
various characteristic lengths of the system, we consider
fluctuations about the homogeneous condensate (\eqcfl).
We parametrize the order
parameter by
$$
\Phi=\phi_0+{1\over\sqrt{6}}({{X+iY}\over 2})
+\bar T^l({{X_l+iY_l}\over 2})
\numbereq\name{\eqdev}
$$
with $X$'s and $Y$'s real, and form linear combinations
of the ordinary electromagnetic 
gauge potental with the eighth component of the
colour gauge potential \ref{\jur}
{M. Alford, J. Berges and K. Rajagopal, \np571, (2000) 269.},
$$
\eqalign{
{\vec V}&={\vec A}^8{\cos\theta}+{\vec{\cal A}}{\sin{\theta}}\cr
{\vec{\cal V}}&=-{\vec A}^8{\sin\theta}+{\vec{\cal A}}{\cos{\theta}}, \cr}
\numbereq\name{\eqelecwk}
$$
where the ``Weinberg angle''  is given by
$$
\tan\theta=-{{2e}\over {{\sqrt 3}g}},
\numbereq\name{\eqweinberg}
$$
Next we expand (\eqrsebb)
to quadratic order in $X$'s $Y$'s $\vec A$'s and ${\cal A}$. We find 
$$
\eqalign{
\Gamma &=\Gamma_{\rm min}+\int d^3{\vec r}\lbrace
{1\over 2}(\vec\nabla\times\vec{\cal V})^2
+{\rm tr}[(\vec\nabla\times\vec W)^2+m_W^2{\vec W}^2]
+{1\over 2}[(\vec\nabla\times\vec Z)^2+m_Z^2{\vec Z}^2]\cr
&+{1\over 2}[(\vec\nabla X)^2+m_H^2 X^2]
+{1\over 2}[(\vec\nabla X^l)(\vec\nabla X^l)+m_H^{\prime2}X^lX^l]
+{1\over 2}(\vec\nabla Y)^2
\rbrace \cr},
\numbereq\name{\eqexpand}
$$
where $\vec W=\sum_{l=1}^{7}\bar T^l\vec W^l$ and
$$
\eqalign{
{\vec W}^l&={\vec A}^l-{1\over m_W}\vec\nabla Y^l \qquad {\rm for}
\quad  l=1, 2, \cdots 7 \cr
{\vec Z}&={\vec V}-{1\over m_Z}\vec\nabla Y^8. \cr}
\numbereq\name{\eqwz}
$$
The masses of the excitations provide us with the
relevant length scales which are the coherence lengths, $\xi$ and $\xi^\prime$ 
defined by:
$$
m_H^2=(b_1+3b_2)\phi_0^2={2\over {\xi}^2}, \qquad 
m_H^{\prime 2}=b_1\phi_0^2={2\over {\xi^\prime}^2},
\numbereq\name{\eqhiggs}
$$
that indicate the distances over which the
di-quark condensate varies
and the magnetic penetration depths, $\delta$ and $\delta^\prime$ by
$$
m_Z^{2}=4g^2{\phi_0^2}{\rm sec}^2\theta={1\over {\delta}^2}, \qquad
m_W^2=m_Z^{2}{\cos^2{\theta}}={1\over {\delta^\prime}^2}.
\numbereq\name{\eqvector}
$$
The excitations that violate (\eqparity)
correspond to the
Goldstone bosons associated with chiral symmetry breaking,
and the $\eta$ particle that becomes massive through the
anomaly.

In the rest of this section, we shall determine
the critical magnetic field of a
homogeneous colour superconductor. 
For simplicity, the diamagnetic response
of the quarks in the normal phase
will be neglected. In the presence of an external magnetic field, the 
thermodynamic function to be minimized is the Gibbs
free energy $\tilde\Gamma$,
which is a Legendre transformation of the
Helmholtz free energy $\Gamma$, i.e.
$$
\tilde\Gamma=\Gamma-\vec H\cdot\int d^3\vec r\vec\nabla\times
\vec{\cal A}.
\numbereq\name{\eqlegdr}
$$
Following the decomposition (\eqelecwk), we write
$$
{\tilde\Gamma}={\tilde\Gamma}_1+{\tilde\Gamma}_2
\numbereq\name{\eqbagger}
$$
with
$$
\eqalign{
{\tilde\Gamma}_1&={\int}d^3{\vec r}\lbrack
{1\over 4}\sum_{l=1}^7F_{ij}^lF_{ij}^l
+{1\over 2}({\vec{\nabla}}\times{\vec V})^2+
4{\rm{tr}}({\vec D}\Phi)^{\dag}({\vec D}\Phi)
+4a{\rm{tr}}\Phi^{\dag}\Phi+b_1{\rm{tr}}(\Phi^{\dag}\Phi)^2\cr 
&+b_2({\rm{tr}}\Phi^{\dag}\Phi)^2
-{\vec H}\cdot{\nabla}{\times}{\vec V}{\sin\theta} \rbrack \cr
{\tilde\Gamma}_2&={\int}d^3{\vec r} \lbrack{1\over 2}({\vec\nabla}\times
{\vec{\cal V}})^2-{\vec H}\cdot{\nabla}{\times}{\vec {\cal V}}
\cos\theta \rbrack \cr}
\numbereq\name{\eqriot}
$$

The minimization of ${\tilde\Gamma}_2$ gives rise to 
$$
{\vec\nabla}\times\vec{\cal V}=\vec H\cos\theta,
\numbereq\name{\eqnormal}
$$
and no distinction will be made between
the super-phase and the normal phase.
In the remaining of the section we shall concentrate
solely on ${\tilde\Gamma}_1$.

A homogeneous super-phase is characterized by a perfect Meissner effect, 
$\vec \nabla\times{\vec{\cal V}}=0$ and the
CFL condensate $\Phi=\phi_0$, which gives rise to 
the minimum of $\tilde\Gamma$,
$\tilde\Gamma_s=\Gamma_{\rm min.}$ given by (\eqfmin).
$\tilde\Gamma_1$ for a
homogeneous normal phase is minimized
by ${\vec\nabla}\times\vec{\cal V}=\vec H\sin\theta$
and the minimum reads $\tilde\Gamma_n=-\Omega{1\over 2}H^2\sin^2\theta$. 
The critical magnetic field $H=H_c$ is determined 
from the equilibrium condition
$$ 
\tilde\Gamma_n=\tilde\Gamma_s=\Gamma_{\rm min}=
-\Omega{12a^2\over b_1+3b_2}
\numbereq\name{\eqequil},
$$
which leads to
$$
H_c=2\sqrt{6a^2\over b_1+3b_2}|{\rm csc}\theta|.
\numbereq\name{\eqcritical}
$$
Substituting the parameters (\eqcoefffb) which were calculated
using weak coupling approximation into (\eqequil)
yields
$$
H_c=4{\sqrt{3\over {14{\zeta}(3)}}}{\mu}^2({{k_BT_c}\over
{\mu}})(1-{T\over T_c})|{\csc{\theta}}|.
\numbereq\name{\eqcrespo}
$$
Extrapolating this expression to a realistic value for
the quark chemical potential, for example $\mu=400$MeV,
and using the one-loop running coupling constant
$$
g^2={{12\pi^2}\over {({11\over 2}N_c-N_f){\ln{{\mu}\over
{\Lambda}}}}}
\numbereq\name{\eqzaneti}
$$
with $N_c=N_f=3$ and $\Lambda=200$MeV, we find
$$
\theta \cong -5.6{^\circ}
\numbereq\name{\eqdida}
$$
and
$$
H_c \cong 1.47 \times 10^{20}({{k_BT_c}\over
{\mu}})(1-{T_c\over T})\quad {\rm Gauss},
\numbereq\name{\eqfero}
$$
which is much higher than the typical magnetic field inside a neutron star. 
This estimation of course depends on the validity of
the extrapolation of the weak coupling formulas to
realistic values of the chemical potential.

\newsection The Domain Wall Problem.%

Let's consider an interface between the normal phase and
the colour superconducting phase, for example
the $y-z$ plane. The equilibrium in the bulk is maintained by 
a uniform external magnetic field of critical magnitude,
$\vec H=H_c\hat\zeta$.
All quantities then in both phases will depend solely on $x$. It
is evident from symmetry that the gauge field lies in one plane,
let this be the $x-y$ plane, so that
${\vec V}=V(x){\hat y}$ and ${\nabla}\times{\vec V}=
{dV\over dx}{\hat\zeta}$.
The boundary conditions on the Ginzburg-Landau equations
in the problem that we are considering
(corresponding to the normal and colour superconducting
phases as $x \mapsto -\infty$ and $x \mapsto \infty$) are
$$
\eqalign{
&\phi \mapsto 0, \quad \chi \mapsto 0, \quad {{dV}\over dx} \mapsto
H_c{\sin\theta} \quad {\rm at} \quad x \mapsto -\infty \cr
&\phi \mapsto \phi_0, \quad \chi \mapsto {\sqrt 2}\phi_0
, \quad {dV\over dx} \mapsto 0 \quad {\rm at} \quad x \mapsto \infty \cr}
\numbereq\name{\eqerionb}
$$
The surface energy $\sigma$ is defined as 
$$
\sigma={{\tilde\Gamma}_1-{\tilde\Gamma}_s\over\hbox{ Area of $yz$ plane }}
\numbereq\name{\eqdafy}
$$
where the Gibbs free energy of a homogeneous superphase, $\tilde\Gamma_s$ 
is given by (\eqequil), and it is equal to that of the normal phase 
in the presence of $H=H_c$.

The minimization of $\tilde\Gamma_1$ generates a set of coupled 
nonlinear differential
equations-the Ginzburg-Landau equations-subject
to the boundary conditions (\eqerionb).
Now we divide the field variables in equation (\eqrioh)
into two groups, the
first group contains $\Phi_0, \Phi^8, {\vec A}^8, {\vec{\cal A}}$,
while the second one  includes $\Phi_l, {\vec A}^l$ with
$l=1, 2, \cdots 7$.
A closer inspection of the structure of
${\vec D}\Phi$ reveals the absence of terms in the free energy
(\eqrseaa) that are linear in the variables of the
second group. Therefore a solution of the equations of motion
exists in which only the fields of the first group acquire
non-zero values. Since this {\it ansatz} implements the maximum symmetry 
allowed by the boundary conditions, we expect that it includes the solution 
that minimizes the free energy of the domain wall.

By setting $\Phi_l={\vec A}^l=0$ for $l=1, 2, \cdots 7$
and transforming the
remaining variables
$$
\phi=\Phi_0+{1\over {\sqrt 3}}\Phi^8, \qquad
\chi={\sqrt 2}(\Phi_0-{1\over {2\sqrt 3}}\Phi^8),
\numbereq\name{\eqwasin}
$$
we arrive
at the following expression for the Ginzburg-Landau
free energy functional (\eqrseaa)
$$
\eqalign{
\tilde\Gamma_{1}=\int d^3{\vec r} \Big[ &{1\over 2}({\vec\nabla}\times
{\vec V})^2+4|({\vec\nabla}-i{g\over {\sqrt 3}}{\sec\theta}
{\vec V})\phi|^2
+4|({\vec\nabla}+i{g\over 2{\sqrt 3}}{\sec\theta}{\vec V})\chi|^2 \cr
&+4a(|{\phi}|^2+|{\chi}|^2)+b_1(|{\phi}|^4+{1\over 2}|{\chi}|^4)
+b_2(|{\phi}|^2+|{\chi}|^2)^2)-H_c{\hat\zeta}\cdot({\vec\nabla}\times {\vec V})
{\sin\theta} \Big]. \cr}
\numbereq\name{\eqgrop}
$$
The condition for the colour-flavour locking ( that the
$3\times 3$ matrix $\Phi$ is proportional to a unitary
matrix ) amounts to $|{\chi}|^2=2|{\phi}|^2$.
Consequently, we shall use instead of the variable $x$ and the
functions $V$, $\phi$
and $\chi$ the dimensionless quantities
$$
s={\delta}x, \quad \phi={\phi}_0u,
\quad \chi={\sqrt 2}{\phi}_0v, \quad
{V}=-{\sqrt{-3a}\over g}A{\cos\theta}.
\numbereq\name{\eqrewe}
$$
The boundary conditions in terms of the new dimensionless
quantities become
$$
\eqalign{
&u \mapsto 0, \quad v \mapsto 0, \quad A^{\prime} \mapsto 1 \quad
{\rm at} \quad s \mapsto -\infty \cr
&u \mapsto 1, \quad v \mapsto 1
, \quad A^{\prime} \mapsto 0 \quad
{\rm at} \quad s \mapsto \infty \cr}
\numbereq\name{\eqerionb}
$$
where prime indicates differentiation with respect to $s$.
The surface energy per unit area in terms of the dimensionless
quantities becomes
$$
\eqalign{
\sigma={6a^2\over b}{\int^{\infty}_{-\infty}}ds&\lbrace
{1\over 2}(A^{\prime}-1)^2+{1\over 3{\kappa}^2}(u^{{\prime}2}
+2v^{{\prime}2})+{1\over 6}A^2(2u^2+v^2)-{1\over 3}(u^2+2v^2)\cr
&+{1\over 18}(2u^4+2u^2v^2+5v^4)+{1\over 18}\rho(u^2-v^2)^2\rbrace\cr}
\numbereq\name{\eqafran}
$$
where
$$
\kappa={\delta\over\xi}
\numbereq\name{\eqxoutos}
$$
is the Ginzburg-Landau parameter, $b={1\over 4}(b_1+3b_2)$,
and
$$
\rho={{b_1-3b_2}\over {b_1+3b_2}}
\numbereq\name{\eqerdogan}
$$
is another dimensionless parameter.
The weak coupling calculation fixes  $\rho$
to be equal to $-{1\over 2}$.
The Ginzburg-Landau equations which determine the
profile of the colour condensate
$u$, $v$ and the magnetic field in the colour superconductor
are determined by minimizing the surface energy with respect to functions $u$, $v$ and $A$
$$
\eqalign{
&-A^{\prime\prime}+{1\over 3}A(2u^2+v^2)=0, \cr
&-{1\over {\kappa}^2}u^{\prime\prime}+(A^2-1)u
+{1\over 3}(2u^2+v^2)u+{1\over 3}
\rho(u^2-v^2)u=0, \cr
&-{1\over {\kappa}^2}v^{\prime\prime}
+{1\over 4}(A^2-4)v+{1\over 6}(u^2+5v^2)v
-{1\over 6}\rho(u^2-v^2)v=0. \cr}
\numbereq\name{\eqfrion}
$$
In the presence of a non-zero $A$, as it is required by the boundary
conditions, this set of equations does not admit a solution in which
$u=v$ everywhere. Stated differently, in the presence of an
inhomogeneity and a non-zero gauge potential
the unlocked
condensate-the octet-has to acquire a non-zero value somewhere.
This situation is analogous to the Ginzburg-Landau theory of a
cuprate superconductor \ref{\ting}{Y. Ren, J.H. Xu and C.S. Ting
\prl74, 3680 (1995).}, where the $s$-wave condensate
becomes non-zero in the vicinity of a vortex filament while the $d$-wave 
condensate dominates in the bulk.
It is easily verified that equations (\eqfrion) have the
first integral
$$
\eqalign{
{1\over 2}A^{{\prime}2}+{1\over 3{\kappa}^2}(u^{{\prime}2}
+2v^{{\prime}2})-{1\over 6}A^2(2u^2+v^2)+{1\over 3}(u^2+2v^2)
&-{1\over 18}(2u^4+2u^2v^2+5v^4)\cr
&-{1\over 18}\rho(u^2-v^2)^2={1\over 2} \cr},
\numbereq\name{\eqdeft}
$$
which implies, according to the boundary conditions (\eqerionb), that 
$$
A(\infty)=0.
\numbereq\name{\eqarlim}
$$

Let $\sigma_{\rm min}$ be the minimum surface energy. It is
obtained from the substitution of the field variables
$u$, $v$, $A$ that satisfy the GL equations (\eqfrion)
into (\eqafran),
and is a function of the parameters $\kappa$ and $\rho$. We proceed now 
to prove the following lemma

\noindent
{\it Lemma}:
$$
\Big({\partial\sigma_{\rm min}\over\partial\kappa}\Big)_\rho\le 0;
\numbereq\name{\eqlemmaa}
$$
$$
\Big({\partial\sigma_{\rm min}\over\partial\rho}\Big)_\kappa\ge 0;
\numbereq\name{\eqlemmab}
$$

\noindent
{\it Proof}:

According to (\eqafran),
$$
\eqalign{
\Big({\partial\sigma_{\rm min}\over\partial\kappa}\Big)_\rho&=
-{4a^2\over b{\kappa}^3}\int_{-\infty}^\infty
ds(u^{\prime 2}+2v^{\prime 2})\cr
&+\int_{-\infty}^\infty ds\Big[{\delta\sigma_{\rm min}\over\delta A(s)}
\Big({\partial A(s)\over\partial\kappa}\Big)_\rho
+{\delta\sigma_{\rm min}\over\delta u(s)}
\Big({\partial u(s)\over\partial\kappa}\Big)_\rho
+{\delta\sigma_{\rm min}\over\delta v(s)}
\Big({\partial v(s)\over\partial\kappa}\Big)_\rho\Big]\cr}
\numbereq\name{\eqlemmapf1}
$$ 
The second term vanishes because of the equation of motion
(\eqfrion) and the boundary conditions (\eqerionb), (\eqarlim).
Consequently, we find that
$$
\Big({\partial\sigma_{\rm min}\over\partial\kappa}\Big)_\rho
=-{4a^2\over b{\kappa}^3}\int_{-\infty}^\infty
ds(u^{\prime 2}+2v^{\prime 2})\le 0.
\numbereq\name{\eqlemmapf2}
$$
Similarly
$$
\Big({\partial\sigma_{\rm min}\over\partial\rho}\Big)_\kappa={a^2\over 3b}
\int_{-\infty}^\infty ds(u^2-v^2)^2\ge 0.
\numbereq\name{\eqlemmapf3}
$$
The lemma is then proved.

Furthermore lets denote the critical value of the
Ginzburg-Landau parameter
for which $\sigma_{\rm min}$ vanishes by
$\kappa_c(\rho)$. We shall prove then the following theorem

\noindent
{\it Theorem}:
$$
\kappa_c(\rho)\le {1\over\sqrt{2}}
\numbereq\name{\eqtheorema}
$$
and
$$
{d\kappa_c\over d\rho}\ge 0.
\numbereq\name{\eqtheoremb}
$$

\noindent
{\it Proof}:

Consider a special field configuration $u=v$, which
satisfies the
boundary conditions (\eqerionb) but not
the Ginzburg-Landau equations (\eqfrion)
with $A\neq 0$. Therefore
$$
\sigma_{\rm min.}\leq{6a^2\over b}\int_{-\infty}^\infty ds
\Big[{1\over 2}(A^\prime-1)^2
+{1\over\kappa^2}u^{\prime 2}
+{1\over 2}A^2u^2-u^2+{1\over 2}u^4\Big]\equiv\bar\sigma
\numbereq\name{\eqtrialb}
$$
The minimization of $\bar\sigma$ with respect to $u$ and $A$
yields the set of differential equations, 
$$
\eqalign{
&-A^{\prime\prime}+Vu^2=0, \cr
&-{1\over {\kappa}^2}u^{\prime\prime}+{1\over 2}(A^2-2)u+u^3=0, \cr}
\numbereq\name{\eqtriala}
$$
subject to the boundary conditions (\eqerionb).
We recognize that (\eqtrialb) and (\eqtriala) correspond exactly
to the domain wall problem for
an ordinary superconductor analyzed by Ginzburg and Landau
in their original
work. It follows then that $\bar\sigma=0$ at
$\kappa={1\over\sqrt{2}}$. On the other hand, this field
configuration does not satisfy
the equations of motion (\eqafran) and therefore
$\sigma_{\rm min}\le 0$ at $\kappa={1\over\sqrt{2}}$. Following
(\eqlemmaa) of the lemma we establish the first equation of the theorem.  

The second equation of the theorem follows from the identity 
$$
\Big({\partial\kappa\over \partial\rho}\Big)_{\sigma_{\rm min}}
=-{\Big({\partial\sigma_{\rm min}\over\partial\rho}\Big)_\kappa
\over\Big({\partial\sigma_{\rm min}\over\partial\kappa}\Big)_\rho}
\numbereq\name{\eqidty}
$$
and the lemma. The theorem is then proved.

The solution to the Ginzburg-Landau equations (\eqfrion) and the 
critical value of the Ginzburg-Landau parameter $\kappa$ for 
various values of $\rho\in(-1,\infty)$ were also determined numerically. The 
continuous range $s\in(-\infty,\infty)$ was replaced by a finite 
lattice with
$$
s_n=-L+n\epsilon
\numbereq\name{\eqsite}
$$
where $n=0,1,2,...,N+1$ and $(N+1)\epsilon=2L$. A good approximation 
amounts to $\epsilon$ and $L$ that satisfy
$\epsilon<<{\rm min}
(1,1/\kappa)$ and $L>>{\rm max}(1,1/\kappa)$. The field variables
$u_n$, $v_n$, and $A_n$ are assigned to each site and their derivatives 
can be approximated by
$u_n^\prime=(u_{n+1}-u_n)/\epsilon$, $v_n^\prime=(v_{n+1}-v_n)/\epsilon$,
and $A_n^\prime=(A_{n+1}-A_n)/\epsilon$. The surface energy is 
then written
$$
\eqalign{
\sigma&={6a^2\over b}\lbrace\epsilon\sum_{n=0}^N\Big[
{1\over 2}(A_n^\prime-1)^2
+{1\over 3\kappa^2}(u_n^{\prime2}+2v_n^{\prime2})\Big]\cr
&+\sum_{n=0}^{N+1}\epsilon_n\Big[{1\over 6}A_n^2(2u_n^2+v_n^2)
-{1\over 3}(u_n^2+2v_n^2)+{1\over 18}(2u_n^4+2u_n^2v_n^2+5v_n^4)
+{1\over 18}\rho(u_n^2-v_n^2)^2\Big]\rbrace \cr}
\numbereq\name{\eqdiscrete}
$$
with $\epsilon_0=\epsilon_{N+1}={1\over 2}\epsilon$ and
$\epsilon_n=\epsilon$ for $n=1, \cdots N$.
The multivariable function (\eqdiscrete) is then minimized by iteration
subject to the conditions $u_0=v_0=0$, $u_{N+1}=v_{N+1}=1$, 
$A_0^\prime=1$ and $A_{N+1}=0$.

The critical GL parameter, $\kappa_c$ versus 
$\rho$ is plotted in Fig.1. The critical value of GL parameter in the weak 
coupling approximation
($\rho=-{1\over 2}$), is $\kappa_c=0.589$.
The two components of the di-quark condensate,
the colour-flavour locked and the colour-flavour octet,
and the corresponding magnetic field, that
correspond to the solution with $\rho=-{1\over 2}$ and
$\kappa=\kappa_c$ are plotted in Fig. 2. The numerical results are consistent 
with the inequalities (\eqtheorema) and (\eqtheoremb) that
we derived analytically.

Substituting the weak coupling expression of the parameters,
the coherence length reads
$$
\xi^2={{7{\zeta}(3)}\over {48{\pi}^2}}\Big({1\over {k_BT_c}}\Big)^2
{T_c\over {T_c-T}},
\numbereq\name{\eqer}
$$
and the penetration depth
$$
\delta^2={{3\pi}\over {2{\alpha}_s{\mu}^2}}{T_c\over {T_c-T}}
{\cos^2\theta}
\numbereq\name{\eqre}
$$
with $\alpha_s={g^2\over 4\pi}$.
The Ginzburg-Landau parameter is given then by
$$
\kappa={{\delta^2}\over {\xi^2}}=
\sqrt{{72\pi^3}\over {7{\zeta}(3){\alpha}_s}}
{{k_BT_c}\over {\mu}}{\cos{\theta}}\cong {16.3\over {\sqrt{\alpha_s}}}
{{k_B}T_c\over {\mu}}.
\numbereq\name{\eqdani}
$$
If we consider a realistic value for the chemical potential,
for example $\mu=400$MeV, it follows from 
(\eqpenecsc) that the colour superconductor will be of type I if 
$k_BT_c < 14$MeV and of type II otherwise.
Furthermore if we extrapolate the weak coupling 
formula for $T_c$ \ref{\son}{D. T. Son
\pr59, 094019 (1999).}, \ref{\schafer}{T. Schafer and F. Wilczek,
\pr60, 114033 (1999).},
\ref{\pisa}{
R. D. Pisarski, and D. H. Rischke, \pr61, 051501 (2000);
\pr61, 074017 (2000).},
\ref{\brown}{W. Brown, J. T. Liu and H-c. Ren, \pr61,
114012 (2000); \pr62, 054016.}, \ref{\wang}{Q. Wang and D. H.
Rischke, \pr65, (2002) 054005.},
which is valid
at asymptotic densities, to $\mu=400$MeV we find $k_BT_c=3.5$MeV and 
the colour superconductor is of type I. Of course
if nonperturbative effects 
raise $T_c$ significantly, for example by one order of 
magnitude, the colour superconductor could be of type II.

\newsection Concluding Remarks.%

In this paper we have applied the Ginzburg-Landau theory to the
problem of a domain wall which separates the normal phase
from the colour superconducting
phase of dense quark matter. The main purpose of this
work was to establish a criterion that determines whether
the colour superconductor is type I or type II
with respect to an external magnetic field. We initially
derived the Ginzburg-Landau equations of motion for the
gauge fields and the order parameter by minimizing the
surface energy of the interface between the two phases.
The condition that the surface energy is a minimum is
equivalent to the requirement that the normal and the
colour superconducting phases are in stable equilibrium
with each other. This equilibrium is maintained by
an external magnetic field. Using symmetry arguments
we were able to simplify the original set of equations
and rewrite them
in terms of only two components of the
order parameter and one gauge potential. The equations also
contain two dimensionless
parameters: the Ginzburg-Landau parameter $\kappa$
and $\rho$.
Therefore the problem is more complicated than the problem
of the metallic superconductor which
was addressed by Ginzburg and Landau.
Although we were unable to derive an analytical result
analogous to the Ginzburg-Landau criterion (\eqpene), we 
derived  several rigorous
inequalities about the critical parameter $\kappa_c$, the
value for which the surface energy vanishes,
and its derivative with respect to $\rho$.
Those analytical results are
supported by a numerical solution of the
Ginzburg-Landau equations. By extrapolating the weak coupling
approximation formulas for the parameters
to realistic values of the chemical potential,
for example $\mu=400$MeV, we find that
the colour superconductor is of type I (type II)
if the transition temperature is below(above) 14MeV.

The value of our work is mainly theoretical. It addresses
the broken $U(1)$ gauge field onto which the electromagnetic
field has a small projection.
Although we have been refering to
the typical chemical potential of a neutron star, the
implications of our work might be quite limited [\jur].
First of all, there has not been
any evidence yet that colour superconductivity
is realised in the core
of a neutron star. Even if this is the case, the critical magnetic field 
(\eqfero) is too high for the distintion between type I and type II to 
be observed unless the pairing force is strong
enough to reduce the lower critical field by several
orders of magnitude. In addition the 
small mixing angle (\eqdida)
and the dependence of the magnetic response on 
the thickness of the crust of the CSC core make it difficult to observe
the partial magnetic Meissner effect and the
appearance of vortex filaments.

Because of the strength of the QCD coupling
and the ultra-relativistic Fermi
sea, the fluctuations of the gauge field 
could be more significant 
than those in a non-relativistic superconductor
\ref{\love}{D. Bailin and A. Love,
\pl109 501 (1982).}.
These fluctuations contribute to the
Ginzburg-Landau free energy functional an energy
density
$$
{\Delta\Gamma} \sim k_BT_c{\delta}^{-3}.
\numbereq\name{\eqerice}
$$
As the condensate energy (\eqfmin) is proportional
to $(1-{T\over T_c})^2$ while $\delta \sim
{\sqrt{1-{T\over T_c}}}$, the contribution
from the fluctuations will eventually dominate as
$T$ approaches $T_c$ and modify the nature of the
phase transition.

The fluctuation energy is estimated in the appendix within the framework
of the Ginzburg-Landau theory,
where we demonstrate that the inclusion
of this energy modifies the phase transition from
second order to first order. The parameter
$$
t={{\ov T}\over T_c}-1
\numbereq\name{\eqopio}
$$
which describes the deviation of the first order transition
temperature $\ov T$ from the second order one $T_c$, is used
as a measure of the importance of the fluctuations.
We find that
for realistic values of the quark chemical potential,
$\mu=400$MeV and $k_BT_c \ge 13$MeV,
$t \le 0.1$. In that case, the Ginzburg-Landau free
energy functional (\eqrseaa) represents a reasonable approximation.

For $t\sim 1$ the approximation employed to derive the Ginzburg-Landau 
free energy (\eqrseaa) with (\eqcoefffb) from QCD
or from the Nambu-Jona-Lasinio
effective action breaks down. Higher powers of the
order parameter have to be included. It would be very interesting
to develop a systematic weak coupling approximation in this case
and we hope to be able to report progress in the future.

\noindent
\newsection Acknowlegments.

We would like like to thank S. Catto, G. Dunne,
H. Hansson and D. Rischke for useful discussions.
This work
is supported in part by US Department of Energy, contract
number DE-FG02-91ER40651-TASKB.

\newsection Appendix.

In this appendix we shall estimate the effect of
the fluctuations of the gauge field.
A systematic evaluation of the contribution from the
fluctuations to the free
energy requires a resummation of a set of ring diagrams
of QCD at finite temperature.
Here we shall follow
the treatment of Bailin and Love [\love]
and estimate the fluctuation terms
within
the framework of the Ginzburg-Landau theory. Due to
the Debye screening the contributions of the
longitudinal components of the gauge potentials
are supressed, and only
the transeverse components contribute in the static limit.
We expect that our simple-minded
estimation captures the fluctuation terms to the
leading order of the condensate.

Assuming that the small fluctuations are governed by the
quadratic form of the Ginzburg-Landau free energy functional,
the shift of
the free energy due to the Meissner effect is given by
$$
\Delta\Gamma=-k_BT\ln{\int[dW][dZ]e^{-S}\over\int[dW][dZ]
e^{-S_0}}
\numbereq\name{\eqbailin}
$$
where $\vec W$ and $\vec Z$ represent the
transverse degrees of freedom and 
$$
\eqalign{
S&={{\beta}\over 2}\int d^3{\vec r}
[(\vec\nabla\times\vec W)^2+m_W^2\vec W^2
+(\vec\nabla\times\vec Z)^2+m_Z^2\vec Z^2]\cr
S_0&={{\beta}\over 2}\int d^3\vec r[(\vec\nabla\times\vec
W)^2+(\vec\nabla\times\vec Z)^2].\cr}
\numbereq\name{\eqlove}
$$
Completing the gaussian integral, we find that
$$
\eqalign{
\Delta\Gamma=&k_BT\int {d^3\vec k\over (2\pi)^3}\Big[
7\ln\Big(1+{m_W^2\over {k^2}}\Big)
+\ln\Big(1+{m_Z^2\over {k^2}}\Big)\Big]\cr
&=-{k_BT_C\over 6\pi}(7\delta^{\prime -3}+\delta^{-3})
=-{4g^3\over {3\pi}}(7+{\sec^2{\theta}})k_BT_c{|\phi|}^3.\cr}
\numbereq\name{\eqkran}
$$
In the previous calculation, dimensional regularization was employed
to eliminate ultraviolet
divergences. By including this term to
the homogeneous Ginzburg-Landau free energy 
with $\Phi=\phi$, we find 
$$
\Gamma=12a|\phi|^2+12b|\phi|^4
-{4g^3\over {3\pi}}(7+{\sec^2{\theta}})k_BT_c|{\phi}|^3.
\numbereq\name{\eqaust}
$$
The presence of the cubic term will induce a first-order
phase transition at
$\bar T>T_c$ The temperature is determined by the condition that the 
free energy curve $\Gamma$ vs. $|\phi|$ becomes tangent to the $\phi$-axis at 
some $\phi\neq 0$, i.e.
$$
a={16\pi\alpha_s^3\over 81b}(7+{\rm sec}^2\theta)^2(k_BT_C)^2.
\numbereq\name{\eqcond}
$$
Substituting the weak coupling expression for $a$ and $b$ we obtain that
$$
t\equiv {\bar T\over T_C}-1
={49{\zeta^2}(3)(7+{\sec^2{\theta}})^2{\alpha^3}_s\over {139968{\pi}^5}}
({{\mu}\over {k_BT_c}})^2.
\numbereq\name{\eqpoin}
$$
With $\mu=400$MeV, and $k_BT_C>13MeV$, $t<0.1$.
As a result the fluctuations
are not signigicant and the Ginzburg-Landau energy functional
(\eqrseaa) represents a reasonable approximation.

\immediate\closeout1
\bigbreak\bigskip

\line{\twelvebf References. \hfil}
\nobreak\medskip\vskip\parskip

\input refs

\vskip 1in

\input epsf
\epsfxsize 3truein
\centerline{\epsffile{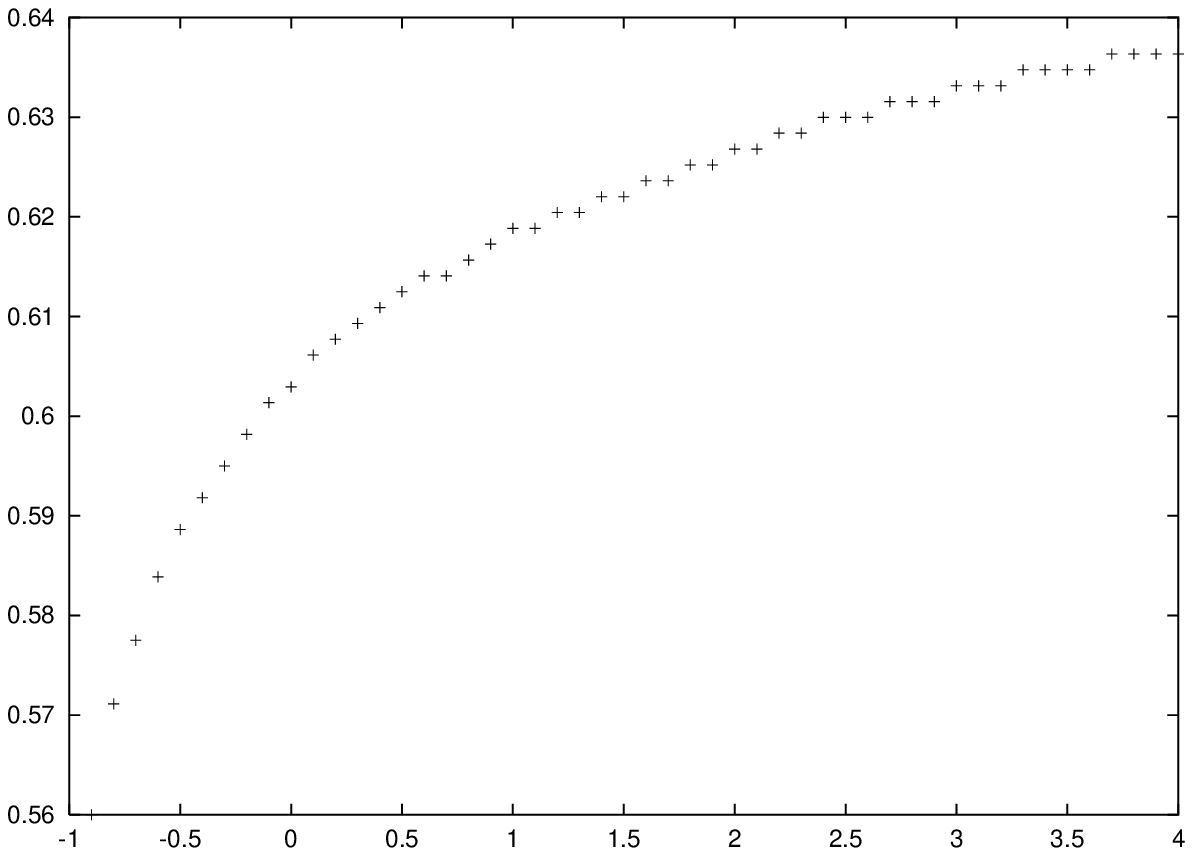}}
\bigskip
\noindent
Figure 1. The critical Ginzburg-Landau
parameter $\kappa_c$ vs. the parameter $\rho$.
\bigskip

\vskip 2in

\input epsf
\epsfxsize 3truein
\centerline{\epsffile{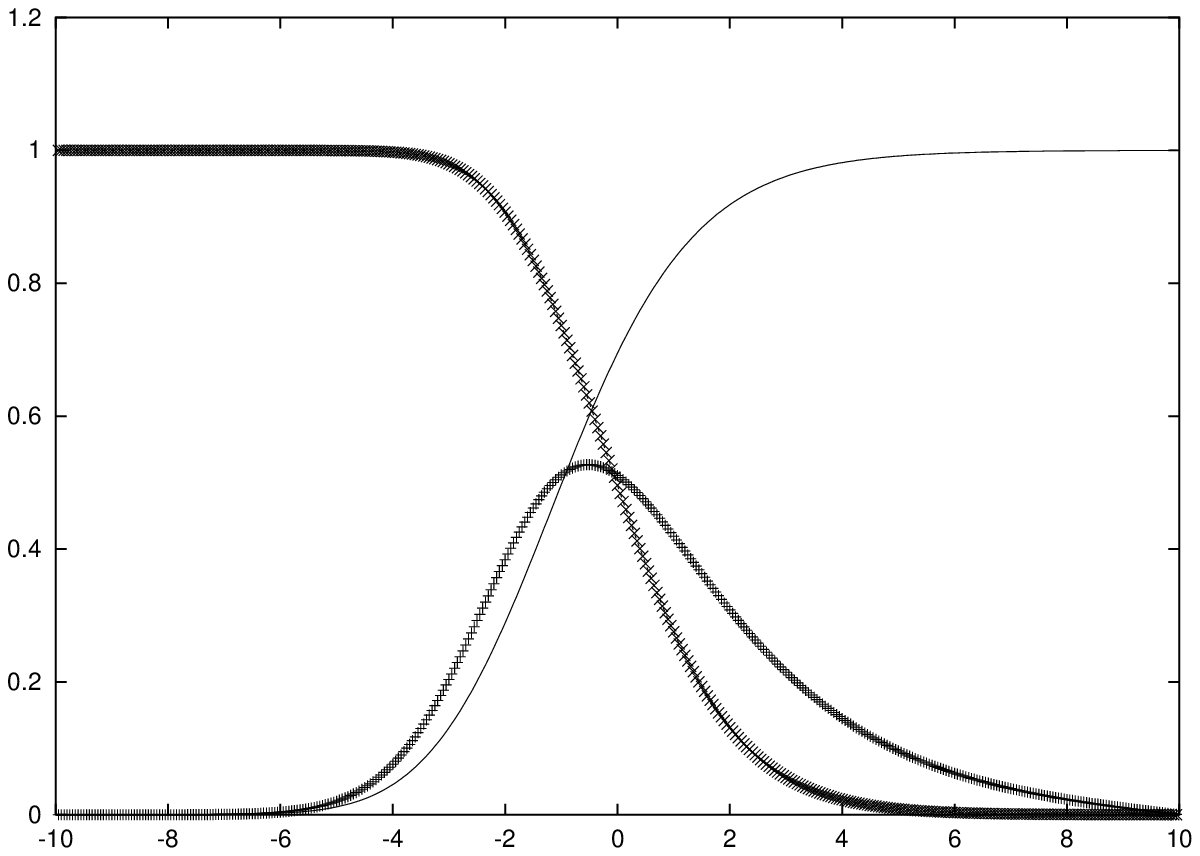}}
\bigskip
\noindent
Figure 2. The solutions to the Ginzburg-Landau equations (\eqfrion)
for $\rho=-{1\over 2}$
and $\kappa=0.589$. The solid line represents the
colour-flavour locked component, 
${1\over 3}(u+2v)$, the dashed line represents the
colour-flavour octet component,
${2\over\sqrt{3}}(v-u)$ and the dotted line the
colour-magnetic field, $A^\prime$.
\bigskip

\vfil\end

\bye